\documentclass{iopart}
\usepackage{iopams}  
\usepackage[latin1]{inputenc}
\usepackage{fancyhdr}
\usepackage{graphicx}
\usepackage{epstopdf}

\usepackage{pstricks}
\usepackage{tocvsec2}

\usepackage{color}

\def\beq{\begin{equation}}
\def\be{\begin{equation}}
\def\ee{\end{equation}}
\def\bes{\begin{eqnarray}}
\def\ees{\end{eqnarray}}

\def\bra{\langle}

\def\f{\frac}

\def\pp{\partial}

\def\arr{\rightarrow}

\def\L{\mathcal{L}}

\begin{document}
\maxtocdepth{subsection}

\title{Diffusion in Curved Spacetimes}

\author{Matteo Smerlak}
\address{Max Planck Institute for Gravitational Physics (Albert Einstein Institute)\\ Am M\"uhlenberg 1, D-14476 Golm, Germany}
\ead{msmerlak@aei.mpg.de}
\date{\small\today}

\begin{abstract}
Using simple kinematical arguments, we derive the Fokker-Planck equation for diffusion processes in curved spacetimes. In the case of Brownian motion, it coincides with Eckart's relativistic heat equation (albeit in a simpler form), and therefore provides a microscopic justification for his phenomenological heat-flux ansatz. Furthermore, we obtain the small-time asymptotic expansion of the mean square displacement of Brownian motion in static spacetimes. Beyond general relativity itself, this result has potential applications in analogue gravitational systems.




\end{abstract}

\pacs{05.10.Gg, 05.60.Cd, 04.90.+e}

\submitto{\NJP}



\section{Introduction}


\subsection{New laws from old ones}
If general relativity is ``probably the most beautiful of all existing physical theories'' \cite{Landau1975}, it is certainly thanks of its geometric character, which reduces the dynamics of test bodies in a gravitational field to pure kinematics.\footnote{``Kinematics'' has several, inconsistent, meanings in the physics literature. Here, we use this term to intend the description of a phenomenon purely in terms of \emph{space} and \emph{time}.} This feature makes it an unprecedented \emph{heuristic machine}: to uncover the effect of gravity on a given physical phenomenon, consider the old law which describes it \emph{in the absence gravity}, phrase it in kinematical terms (times lapses and distance intervals), replace ``time'' by ``proper time'' and ``distance'' by ``proper distance'', and read off the new law \emph{in the presence of gravity}. For instance, the Fermat principle states that, in the absence of gravity, light rays follow the paths which \emph{extremize} \emph{time}. Then general relativity immediately generates a new law from Fermat's principle: in the presence of gravity, light rays follow the paths which extremize \emph{proper time}. The bending of light in the presence of spacetime curvature follows immediately from this new law.  

It is interesting to note that the generative character of general relativity is actually more general than general relativity itself. Indeed, it relies neither on local Lorentz invariance, nor on the absence of a preferred foliation of spacetime, nor on the Einstein equation, nor even on the relationship between stress-energy density and spacetime curvature. This is particularly clear in the example mentioned above: the spacetime curvature responsible for the bending of light could come from a refractive index gradient as well as from the vicinity of a massive star. In other words, whether spacetime curvature is \emph{fundamental} or \emph{effective} is irrelevant to the generation of new law from old ones: all that matters is that \emph{a curved spacetime, seen through geodesic coordinates, appears flat}. 

This circumstance is all the more important that a growing number of condensed-matter systems are now understood to behave as effective, or \emph{analogue}, spacetimes. Besides Gordon's refractive optical medium, one can mention Unruh's dumb hole (a supersonic fluid flow), but also corrugated graphene sheets, Bose-Einstein condensates, slow light systems, superfluids, metamaterials, etc. (See \cite{Barcelo2011} for an updated review of analogue gravity.) The fruitfulness of this connection between condensed-matter physics and general relativity goes both ways: gravitational analogues provide valuable model systems to emulate otherwise out-of-reach relativistic phenomena \cite{Unruh1981}; \emph{vice versa}, the geometric setup of general relativity sheds a new light on venerable fields such as optics \cite{Leonhardt2010}, hydrodynamics \cite{Jannes2011} -- or diffusion phenomena, as this paper intends to demonstrate.


\subsection{The Tolman-Ehrenfest law of thermal equilibrium}

That the ``new law from old ones'' principle does not restrict to the realm of mechanics, or electrodynamics, is demonstrated by the early history of general relativity. Indeed, one the first problems which Einstein analyzed in terms of gravitational redshift was one of \emph{thermodynamics}: the problem of finding the equilibrium temperature distribution $T^{*}$ in a static gravitational field. As early as 1912, that is three years \emph{before} the completion of general relativity, he speculated that, because in a curved spacetime proper time does not run at the same rate in different places, $T^{*}$ should \emph{not} be homogeneous \cite{Einstein1912a}. This remarkable intuition was put on firm ground by Tolman and Ehrenfest, who showed that 
\be\label{tolman}
\chi T^{*}=\textrm{const.},
\ee
where $\chi=(-\xi^{a}\xi_{a})^{1/2}$ is the redshift factor and $\xi^{a}$ a timelike Killing vector \cite{Tolman1930}.  

From our perspective, the Tolman-Ehrenfest relation \eref{tolman} is a prototype of these kinematical laws which follow directly from the geometrical setup of general relativity. In their original paper, however, Tolman and Ehrenfest gave it a complicated \emph{dynamical} proof, which relied both on the Einstein equation and the equation of state of thermal radiation. Several authors later pointed out this anomaly, and proposed more minimalist derivations \cite{Ebert1973,Stachel1984}.  One, due to Rovelli and the author, goes like this \cite{Rovelli:2010kx}:
\begin{itemize}
\item
In the non-relativistic canonical ensemble, the equilibrium temperature $T^{*}$ can be computed as the rate of the modular flow generated by a thermal state $\rho$ (the Hamiltonian flow of $\ln\rho$) with respect to time. 
\item
Hence in a stationary spacetime, $T^{*}$ can be computed \emph{locally} as the rate of the modular flow generated by a thermal state $\rho$ (the Hamiltonian flow of $\ln\rho$) with respect to \emph{proper time}. 
\item
By stationarity, the modular flow generated by $\rho$ is proportional to the Killing flow, hence $T^{*}(\sigma)\propto dt(\sigma)/ds(\sigma)$, where $t(\sigma)$ is the Killing parameter at a spatial point $\sigma$ and $s(\sigma)$ the local proper time. 
\item
By definition, this ratio is the inverse of the redshift factor $\chi(\sigma)$, hence $T^{*}(\sigma)\propto1/\chi(\sigma)$. 
\end{itemize}
In other words, the equilibrium temperature distribution $T^{*}$ is such that the combination $\chi T^{*}$ satisfies the non-relativistic equilibrium criterion. Can this reasoning be extended to non-equilibrium processes such as \emph{heat conduction}, or more general diffusion processes? How far does the ``new laws from old ones'' principle lead us when we leave the equilibrium regime?

\subsection{The stochastic route to diffusion}

As already emphasized, the key step to address this question is to frame the diffusion problem in \emph{kinematical terms}. Unsurprisingly, the most direct route to that effect is the one laid down by Einstein himself, in his 1905 work on Brownian motion \cite{Einstein1905}: write the master equation for a stochastic process, and derive the corresponding Fokker-Planck equation in the diffusive limit. 

In this stochastic approach, the basic concept is that of \emph{transition rates}: the probability that a random walker will jump from one position to another per unit time. These are kinematical in nature (they are expressed in terms of distance and time), and are therefore easily amenable to the ``new laws from old ones'' principle. Precisely by this token, we will derive in this article the curved-spacetime generalization of the Fokker-Planck equation, applicable to any kind of diffusive transport (atomic and molecular diffusion, photon diffusion, thermal and electronic conduction, etc.), in any kind of fundamental or analogue spacetime. In the case of pure Brownian motion and static spacetimes, it reads
\be\label{heat}
\widehat{\xi^{a}}\nabla_{a}(\chi p)=\kappa\Delta(\chi p),
\ee
where $p$ is the probability density of Brownian motion, $\widehat{\xi}^{a}=\xi^{a}/\chi$ the hydrostatic $4$-velocity, $\Delta$ the spatial Laplace-Beltrami operator and $\kappa$ the diffusivity. Note that equation \eref{heat} is nothing but the standard diffusion equation, with $p$ replaced by $\chi p$, as in the Tolman-Ehrenfest relation -- plain and simple. 

Although (to our knowledge) it was never written in this form, equation \eref{heat} is actually well-known in relativistic hydrodynamics: it is the phenomenological heat equation of Eckart \cite{Eckart1940} and Landau and Lifschitz \cite{Landau1987}. Here, it is \emph{derived} from stochastic mechanics, rather than postulated to satisfy the second law of thermodynamics -- just like Einstein \emph{derived} the diffusion equation postulated by Fourier and Fick. 

%
%

%
%

\subsection{On causality}

The notion of ``relativistic Brownian motion'' has been discussed by many authors in the past decades \cite{Dunkel2009}. What is usually meant by this expression is a formulation of Brownian motion that is consistent with special-relativistic \emph{causality}. Our goal here is different: we wish to understand the effect of a non-trivial \emph{spacetime geometry} on a diffusion process. 

As is already apparent from our equation \eref{heat}, which is \emph{parabolic} and hence permits superluminal propagation, we do not attempt to include causality in our framework. Instead, our setup is the following. We consider a fluid flow in spacetime -- the bath within which the stochastic process takes place -- and fix the associated orthogonal foliation of spacetime. In this foliation, each hypersurface is the ``instantaneous space'' relative to fiducial observers comoving with the flow; at each instant $t$, a random walker dragged by the flow can jump from one point to another point of the \emph{same} hypersurface. 

Thus, the sole difference between our approach and the non-relativistic theory is the inclusion of a non-trivial spacetime geometry, viz. the presence of a \emph{gravitational field}. That is to say that we consider the $c\rightarrow\infty$ limit of a fully relativistic theory of Brownian motion, in the same sense in which Newton-Cartan gravity \cite{Friedrichs1927,Cartan1923} is the $c\rightarrow\infty$ limit of general relativity. 

Of course, this acausal approach is at variance with the textbook notion that special relativity \emph{precedes} general relativity. Yet, it is by no means unreasonable. It simply expresses a physical approximation, namely that the relaxation of the bath occurs on time scales much shorter than that of the diffusion process itself. This regime is well-known in the context of kinetic theory, where it is referred to as the ``hydrodynamic limit'', and is also the regime in which Eckart's dissipative relativistic hydrodynamics applies.\footnote{Eckart's theory is often considered ``unacceptable'' because of its acausal character, and its alleged instability \cite{Hiscock1985}. As far as the author can see, this judgement is completely misled: Eckart's heat equation is a perfectly well-behaved parabolic PDE, whose status with respect to a fully relativistic dissipative hydrodynamics is the same as that of Newton-Cartan gravity  with respect to general relativity: an excellent approximation in most physical situations. See \cite{Kostadt2000} for a mathematical argument to this effect.} We refer the reader to \cite{Debbasch1998} for an interesting discussion on why an acausal equation like \eref{heat} is \emph{not} inconsistent with microscopic causality.

\subsection{Results}

Besides deriving the master and Fokker-Planck equations for stochastic processes in curved spacetimes, our results in this paper are the following.

\begin{itemize}
\item
We provide a microscopic justification to Eckart's heat-flux ansatz, and extend it to more general diffusion processes.
\item
We generalize the Tolman-Ehrenfest relation to non-equilibrium stationary states, with arbitrary boundary conditions.
\item
We compute the gravitational corrections to the mean squared displacement of Brownian motion in static isotropic spacetimes.
\end{itemize}
The last item is particularly interesting. In a curved spacetime, the usual scaling law $\langle x^{2}\rangle_{t}\propto t$ holds only in the $t\rightarrow0$ asymptotic limit. At later times, spacetime curvature corrections show up and modify the growth rate of $\langle x^{2}\rangle_{t}$. This suggests that diffusive transport in gravitational analogues could perhaps be \emph{tailored}, by tuning the effective metric coefficients \cite{Smerlak}.

\subsection{Plan of the paper}

%
%
%



The paper is organized as follows. Sec. \ref{prel} consists of preliminaries on the $D$+1 formalism for relativistic hydrodynamics and on the non-relativistic theory of stochastic processes. Our theory of stochastic processes in curved spacetimes is developed in sec. \ref{main}, and the limit case of Brownian motion is studied in sec. \ref{brownian}. In sec. \ref{corrections}, we obtain a small-time asymptotic expansion for the mean squared displacement of Brownian motion in static isotropic spacetimes.  Our conclusion follows in sec. \ref{conclusion}.

\section{Preliminaries}\label{prel}


Throughout this paper, we consider a $(D+1)$ dimensional spacetime with signature $(-++\cdots)$. (We keep $D$ unspecified to include lower dimensional analogue spacetimes in the discussion.) We denote $\nabla$ the spacetime Levi-Civita connection, and $a,b,c ..., i, j, ...$ are abstract indices.

The standard references for general relativity and the $D+1$ formalism are \cite{Misner1973,Wald1984}; stochastic processes and Fokker-Planck equations are exposed in \cite{VanKampen1992,Risken1989}. 

\subsection{The $D$+1 formalism}

Consider a relativistic fluid with velocity $u^a$. Assume its flow is \emph{irrotational}, viz. 
\be
u_{[a}\nabla_bu_{c]}=0.
\ee 
Then, according to the Frobenius theorem, there is a foliation of spacetime by hypersurfaces $\Sigma_t$ orthogonal to $u^a$. Furthermore, the slices $\Sigma_{t}$ are the level sets of a time functions $t:M\rightarrow\mathbb{R}$ such that 
\be
u_a=-N\nabla_at
\ee 
for some non-negative function $N$. The function $N$ is called the \emph{lapse function}, and the slices $\Sigma_{t}$ have the interpretation of ``instantaneous space" relative to observers comoving with the fluid. In the following, we will denote $\sigma$ a flow line of $u^a$ (a ``spatial point''), and $\sigma_{t}$ its intersection with $\Sigma_{t}$. 

The intrinsic geometry of the spatial hypersurfaces $\Sigma_{t}$ is coded by the induced metric 
\be
h_{ab}=g_{ab}+u_au_b,
\ee 
and its associated covariant derivative\footnote{The covariant derivative $D_{a}$ associated to $h_{ab}$ acts on a tensor field $T^{a_{1}\cdots a_{n}}_{\quad\quad b_{1}\cdots b_{m}}$ according to
$$D_{c}T^{a_{1}\cdots a_{n}}_{\quad\quad b_{1}\cdots b_{m}}=h^{\ a_{1}}_{e_{1}}\cdots h_{b_{m}}^{\ d_{m}} h^{\ f}_{c}\nabla_{f} T^{d_{1}\cdots d_{n}}_{\quad\quad e_{1}\cdots e_{m}}.$$} $D_{a}$ and Laplace-Beltrami operator $\Delta$, while their embedding in spacetime is measured by the (symmetric) extrinsic curvature tensor
\be
K_{ab}=\nabla_{a}u_{b}.
\ee
The trace $\theta=K_{a}^{a}=\nabla_{a}u^{a}$ of the extrinsic curvature tensor is called the \emph{expansion} scalar. It measures the fractional rate of change of an infinitesimal volume $\delta V$ about a spatial point along the flow, viz.
\be\label{expansion}
\theta=u^{a}\nabla_{a}\ln \delta V=\f{1}{N}\f{1}{\delta V}\f{d(\delta V)}{dt}.
\ee
The factor $1/N$ above converts the proper time along the flow to the global time coordinate $t$.

A situation of particular interest is the \emph{hydrostatic equilibrium}: the vector $\xi^{a}=\nabla^{a}{t}=-u^{a}/N$ is then \emph{Killing}, i.e. generates timelike isometries. In this context, the lapse function $N$ is usually denoted $\chi$, and called the \emph{redshift factor}. It satisfies $u^{a}\nabla_{a}\chi=0$, and gives the acceleration $a^{b}=u^{c}\nabla_{c}u^{b}$ of the flow by
\be
a^{b}=\nabla^{b}\ln \chi.
\ee
Moreover, the time-time component of the Ricci tensor $E=R_{ab}u^{a}u^{b}$ (sometimes called the Raychaudhuri scalar), is given in this case by
\be
E=D_{b}a^{b}+a_{b}a^{b}.
\ee
In general relativity, this scalar is tightly related to the local mass-energy density, by virtue of the Einstein equation. We will see that $E$ plays an interesting r\^{o} in diffusion phenomena. 



\subsection{Markov processes}\label{reviewmarkov}

Let $\Sigma$ be a Riemannian manifold with metric $h_{ij}$ and covariant derivative $D_{i}$, representing a curved \emph{space}, and denote $\sigma_{t}\in\Sigma$ the instantaneous position of a random walker at time $t$. In the Markovian setup, we assume that $\sigma_{t}$ completely determines its later positions $\sigma_{t'}$ ($t'>t$), according to \emph{transition rates} $\Gamma(\sigma\rightarrow\sigma')$. By definition, these are such that the elementary probability for the walker to jump from a volume $dV(\sigma)$ about $\sigma\in\Sigma$ to a volume $dV(\sigma')$ about $\sigma'\in\Sigma$ in time $dt$ is given by
\be
\Gamma(\sigma\arr\sigma')dV(\sigma)dV(\sigma')dt.
\ee
As a rule, the transition rates are implicit functions of the metric $h_{ij}$.

Let $p_{t}(\sigma)$ denote the probability density that the walker is in neighborhood of $\sigma$ at time $t$, i.e. $\sigma_{t}=\sigma$, and
\be
j_{t}(\sigma\arr\sigma')=p_{t}(\sigma)\Gamma(\sigma\arr\sigma'). 
\ee
the corresponding \emph{probability fluxes}. Balancing the incoming and outgoing fluxes at $\sigma$, we can immediately write the evolution equation for $p_{t}$ as
\begin{equation}
\pp_tp_{t}(\sigma)=\int_\Sigma dV(\sigma')\Big(j_{t}(\sigma'\arr\sigma)-j_{t}(\sigma\arr\sigma')\Big),
\end{equation}
i.e.
\begin{equation}\label{masternonrel}
\pp_tp_{t}(\sigma)=\int_\Sigma dV(\sigma')\Big(p_{t}(\sigma')\Gamma(\sigma'\arr\sigma)-p_{t}(\sigma)\Gamma(\sigma\arr\sigma')\Big),
\end{equation}
where $dV(\sigma_{t})$ is the Riemannian volume element on $\Sigma_{t}$. This integro-differential equation is known \emph{master equation}, and the operator $\mathcal{M}$ such that $\pp_{t}p_{t}=\mathcal{M}p_{t}$ as the \emph{master operator}.

In this stochastic framework, the notion of \emph{equilibrium state} has a clear-cut definition: a steady-state solution $p^{*}$ is an \emph{equilibrium distribution} if the corresponding probability fluxes cancel pairwise, i.e. 
\be
p^*(\sigma)\Gamma(\sigma\arr\sigma')=p^*(\sigma')\Gamma(\sigma'\arr\sigma)
\ee
This condition is known as the \emph{detailed balance condition}.

Under certain regularity conditions for the rates $\Gamma$, one can show that the paths $(\sigma_{t})$ are
discontinuous: for this reason one often speaks of \emph{jump processes} in this case. The situation changes in the limit where the jumps become infinitely frequent and short-ranged (with respect to some relevant coarse-graining scale). Then $\Gamma$ becomes \emph{distributional}, and the master operator $\mathcal{M}$ reduces to its second-order truncation $\L$ in a moment expansion, reading
\be\label{fpoperator}
\mathcal{L}p_{t}=-D_i(w_1^i p_{t})+\f{1}{2}D_{i}D_j(w_2^{ij} p_{t}).
\ee
Here $w_{1}^{i}$ is a vector field on $\Sigma$, the \emph{drift vector}, and $w_{2}^{ij}$ a symmetric and positive-definite rank-$2$ tensor field, the \emph{diffusion tensor}. Note that the transition rates $\Gamma$ are related to $\mathcal{L}$ according to
\be
\Gamma(\sigma'\arr\sigma)=\mathcal{L}\delta(\sigma',\sigma),
\ee
where $\delta$ is the Dirac distribution on $\Sigma$ and $\mathcal{L}$ acts on the $\sigma'$ variable. Stochastic processes described by such Fokker-Planck equations are called \emph{diffusion processes}.

The simplest example of such a diffusion process is \emph{Brownian motion}, for which (by definition) $w_{1}^{j}=0$ and $w_{2}^{ij}=2\kappa h^{ij}$ for some positive constant $\kappa$. The corresponding Fokker-Planck equation $\pp_{t}p_{t}=\mathcal{L}p_{t}$ is the well-known diffusion equation
\be
\pp_{t}p=\kappa \Delta p_{t}.
\ee

\section{Master and Fokker-Planck equations in curved spacetimes}\label{main}

In this section we describe the curved spacetime generalization of the master and Fokker-Planck equations for Markov processes.

\subsection{Markovian setup}

Consider a Markov process defined by stationary transition rates $\Gamma(\sigma'\arr\sigma)$, depending parametrically on a Riemannian metric $h_{ab}$. In the case of Brownian motion, for instance, $\Gamma(\sigma'\arr\sigma)=\kappa\Delta\delta(\sigma,\sigma')$, with $\Delta$ the Laplace-Beltrami operator associated to $h_{ab}$. 

Following the general ``new law from old ones'' ansatz, we take this process as defining the \emph{instantaneous} dynamics of a random walker in spacetime, in \emph{proper time}. In other words, given an irrotational flow $u^{a}$, we consider the associated orthogonal foliation $(\Sigma_{t})$, evaluate $\Gamma$ on the induced metric $h_{ab}$,\footnote{If spacetime is not static, this makes the transition rates implicit functions of time.} and assume that the probability that a random walker carried by the flow $u^{a}$ will jump from the position $\sigma_t$ to the position $\sigma'_t$ in proper time $ds(\sigma_t)$ is given that
\be\label{ratescurved}
\Gamma(\sigma_t\rightarrow\sigma'_t)dV(\sigma_{t})dV(\sigma_{t}')ds(\sigma_t),
\ee
where $ds(\sigma_{t})$ is the proper time along $\sigma$.

\subsection{Master equation}
Now, to write the corresponding probability equation, which is necessarily \emph{global}, we must convert the proper time $s(\sigma_{t})$ in \eref{ratescurved} into the time coordinate $t$. This is achieved thanks to the lapse function $N$, as
 \be
 ds(\sigma_t)=N(\sigma_t)dt.
 \ee
 Hence, we can rewrite \eref{ratescurved} as
\be
\Gamma(\sigma_t\rightarrow\sigma'_t)dV(\sigma_{t})dV(\sigma_{t}')N(\sigma_t)dt.
\ee
Denoting $p(\sigma_{t})$ the probability density of the stochastic process, the probability flux is therefore
\be\label{fluxes}
j(\sigma_{t}\arr\sigma_{t}')=N(\sigma_{t})p(\sigma_{t})\Gamma(\sigma_{t}\arr\sigma_{t}').
\ee
This expression is physically intuitive: \emph{where proper time runs faster (high $N$), the walker jumps more frequently (high $j$)}. 

From this simple argument, we get that, if $\mathcal{M}$ is the master operator associated to the rates $\Gamma$, the right-hand side of the curved-spacetime master equation should be $\mathcal{M}(Np)$, i.e.
\be
\int_{\Sigma_{t}} dV(\sigma'_{t})\Big(N(\sigma'_{t})p_{t}(\sigma'_{t})\Gamma(\sigma'_{t}\arr\sigma_{t})-N(\sigma_{t})p_{t}(\sigma)\Gamma(\sigma_{t}\arr\sigma'_{t})\Big).\ee

A moment of reflection shows that the left-hand side of the master equation should also be modified in a curved spacetime. Indeed, recall that in a curved spacetime, the time-variation of an integrated density does not coincide with the integral of the time-derivative of the density: it $V_{t}$ is a region in $\Sigma_{t}$, then
\be
\f{d}{dt}\int_{V_{t}}dV(\sigma_{t}) p_{t}(\sigma_{t})\neq\int_{V_{t}}dV(\sigma_{t}) \pp_{t}p_{t}(\sigma_{t}).
\ee
This is due to the fact that the volume element $dV(\sigma_{t})$ itself depends on time. The correct formula follows from the relationship \eref{expansion} defining the expansion scalar, and reads
\be
\f{d}{dt}\int_{V_{t}}dV(\sigma_{t}) p_{t}(\sigma_{t})=\int_{V_{t}}dV(\sigma_{t}) \Big(\pp_{t}p_{t}(\sigma_{t})+N\theta p_{t}\Big).
\ee
Shrinking the volume $V_{t}$ down to zero, we thus find that the left-hand side of the master equation should be $\pp_{t}p+N\theta p$ instead of $\pp_{t}p_{t}$.

Combining both insights, we find that the master equation in a curved spacetime with lapse $N$ and expansion $\theta$ is
\be\label{master}
\pp_{t}p+N\theta p=\mathcal{M}(Np).
\ee
It is easy to check that this equation conserves the total probability $\int_{\Sigma_{t}}dV(\sigma_{t})p_{t}(\sigma_{t})$, as it should.

\subsection{Detailed balance condition}

Note that, in the case of static spacetimes ($\theta=0$ and $N=\chi$ is the redshift factor), we can read off from \eref{fluxes} the generalized detailed balance condition: for an equilibrium distribution $p^{*}$, the probability fluxes cancel pairwise if
\be
\Gamma(\sigma'\rightarrow\sigma)\chi(\sigma')p^*(\sigma')=\Gamma(\sigma\rightarrow\sigma')\chi(\sigma)p^*(\sigma).
\ee
Hence, the \emph{product} $\chi p^*$ must satisfy the usual detailed balance condition defined by the rates $\Gamma(\sigma\rightarrow\sigma')$, instead of $p^{*}$ itself, as in the non-relativistic case. This is the stochastic counterpart of the Tolman-Ehrenfest relation \eref{tolman}, where $\chi T^{*}$ itself satisfies the usual homogeneity condition instead of $T^{*}$. 


%

\subsection{Diffusive limit}

Assume from now on that the stochastic process is of diffusive type (or can be approximated by one\footnote{We recommend van Kampen's note \cite{VanKampen} for a discussion of the applicability of this approximation.}) and denote $\mathcal{L}$ the Fokker-Planck operator defined by the rates $\Gamma$, as in \eref{fpoperator}. Then from \eref{master} it follows immediately that the Fokker-Planck equation reads 
\be
\pp_tp+N\theta p=\L(Np),
\ee 
i.e. 
\be\label{fokkerplanck}
\pp_tp+N\theta p=-D_a(w_1^aN p)+\f{1}{2}D_{a}D_b(w_2^{ab}N p)
\ee
where $w_{1}^{a}$ and $w_{2}^{ab}$ are the drift vector and diffusion tensor associated to the rates $\Gamma$, as in sec. \ref{reviewmarkov}.
This is the \emph{curved-spacetime Fokker-Planck equation}. 

Note that \eref{fokkerplanck} can be given a more hydrodynamical flavor, by replacing the unphysical derivative $\pp_t$ by the convective derivative $u^a\nabla_a$, which evolves the probability distribution in proper time rather than in coordinate time; it then becomes
\be\label{fp2}
u^a\nabla_ap+\theta p=-\f{D_a(w_1^aN p)}{N}+\f{1}{2}\f{D_aD_{b}(w_2^{ab}N p)}{N}.
\ee
This equation is the main result of this paper.

\section{The case of Brownian motion}\label{brownian}

In this section we focus on the properties of \emph{Brownian motion} in curved spacetimes. 

\subsection{The general-relativistic diffusion equation}

We saw in sec. \ref{prel} that Brownian motion is characterized among diffusion processes by the vanishing of the drift vector, $w_{1}^{a}=0$, and by $w_{2}^{ab}=\kappa h^{ab}$, with $\kappa$ the diffusivity. The corresponding Fokker-Planck equation is therefore
\be\label{lapseddiffusion}
\pp_{t}p+N\theta p=\kappa\Delta (Np)
\ee
or
\be\label{lapseddiffusion2}
u^{a}\nabla_{a}p+\theta p=\kappa\f{\Delta (Np)}{N}.
\ee
The remainder of this article is concerned with the properties of this \emph{curved-spacetime diffusion equation}.
\subsection{Comments in the hydrostatic case}\label{hydrostatic}

Consider the hydrostatic case, where \eref{lapseddiffusion2} reduces to
\be\label{lapseddiffusionstatic}
u^{a}\nabla_{a}p=\kappa\f{\Delta (\chi p)}{\chi}.
\ee
Several comments can be made about this equation. First, since $u^{a}\nabla_{a}\chi=0$, this equation indeed coincides with \eref{heat}, as announced in the introduction. Second, using the relation $a_b=D_b\log\chi$ between the acceleration of the congruence $a_b$ and the spatial gradient of the redshift factor, the equation \eref{lapseddiffusionstatic} can be reorganized as
\be\label{threeterms}
(\widehat{\xi}^{b}-2\kappa a^b)\nabla_{b}p=\kappa\Delta p+\kappa Ep,
\ee
where $E$ is the Raychaudhuri scalar. In addition to the usual diffusion term $\Delta p$, this equation contains two remarkable terms, which have no analogue in the non-relativistic diffusion equation. 
\begin{itemize}
\item\emph{Drift}. The term $2\kappa a^{b}\nabla_{b}p$ is a \emph{drift term}. Unlike the drift term in the classical Fokker-Planck equation \eref{fpoperator}, it vanishes in the limit $\kappa\arr0$, and is therefore a genuine effect of diffusion. 
\item\emph{Source}. The term $\kappa Ep$, where $E=D_ba^b+a_ba^b$, is a \emph{source term}. It implies that the probability density appears to comoving observers as sourced by ($\kappa$ times) the Raychaudhuri scalar $E$.\footnote{That is \emph{not} to say that the total probability is not conserved; we saw that it is.} 
\end{itemize}
Both terms, which result from the non-homogeneity of $\chi$ in space, can be interpreted as stochastic \emph{gravitational redshift} effects.

\subsection{Derivation of Eckart's constitutive relation}
Another interesting consequence of our stochastic derivation of the diffusion equation in curved spacetimes is the vindication of Eckart's phenomenological constitutive relation for the heat flux in general relativity \cite{Eckart1940}:
 \be q^b=-\kappa(D^bT+Ta^b).\ee
This relation was postulated by Eckart on the basis of thermodynamical arguments, and can be used to write the relativistic heat equation as 
 \be\label{eckartheat} u^{a}\nabla_{a}T+\big(D_b+a_b\big)q^b=0.\ee
 
Consider now the diffusion equation \eref{lapseddiffusionstatic} for the probability density of Brownian motion in a static spacetime, and compare it with Eckart's heat equation \eref{eckartheat}: they are the same. In other words, we have reappraised Eckart's heat equation as a probabilistic equation -- just like Einstein did with Fick's diffusion equation.


\subsection{The non-equilibrium Tolman-Ehrenfest condition}

Another straightforward consequence of the lapsed heat equation is the generalization of the Tolman-Ehrenfest condition to non-equilibrium steady-state solutions. 
Indeed, we see from \eref{lapseddiffusion}-\eref{lapseddiffusion2} that the the steady-state solution $T^{\infty}$ is given by \eref{tolman} only in the absence of an external forcing on the boundary; in general, it satisfies instead
\be
\Delta(\chi T^{\infty})=0.
\ee
Hence, steady-state solutions can be described as $T^{\infty}=\psi/\chi$, where $\psi$ is a \emph{harmonic function}. (Equilibrium distributions corresponding to the case $\psi=\textrm{const}$.) To our knowledge, this characterization of steady-state temperature distributions in static spacetimes was not derived before. 

\section{Corrections to the mean square displacement}\label{corrections}

In this section we compute the gravitational corrections to the mean square displacement of Brownian motion as a function of time. 

\subsection{Assumptions}

To avoid dealing with the drift effect mentioned in sec. \ref{hydrostatic}, we assume from now on that space is radially symmetric about $o$, the origin of the Brownian motion. We also assume that the metric is \emph{quenched}, i.e. evolves at a much slower rate than the diffusion process itself. In this approximation, the lapse function $N$ and spatial geometry $h_{ab}$ are essentially independent of $t$, and the expansion scalar $\theta$ is negligible with respect to the (inverse) diffusion time, hence \eref{lapseddiffusion} reduces to
\be\label{staticdiffusion}
\pp_{t}K_{t}=\kappa\Delta(\chi K_{t}).
\ee 
Hereafter, we shall denote $\Sigma$ the time-independent spatial section, and $\langle T,\phi\rangle$ the pairing between a distribution $T$ and a test function $\phi$ on $\Sigma$. We also assume (without loss of generality) that $N(o)=1$. Finally, we disregard the possible existence of cut loci in $\Sigma$, and effectively restrict our attention to a convex normal neighborhood of $o$, where the (spatial) Riemannian distance $\rho(\sigma)=d(\sigma,o)$ is a smooth function of $\sigma$. 

\subsection{Green function and mean square displacement}

The most significant observable of Brownian motion is the \emph{mean square displacement} (MSD). It is defined as the expected value of the squared distance between the position of the Brownian walker at time $t$ and its initial position:
\be\label{defMSD}
\langle \rho^{2}\rangle_{t}=\langle K_{t},\rho^{2}\rangle.
\ee
Here $K_{t}$ is the Green function (or heat kernel) of the diffusion equation \eref{staticdiffusion}, namely the solution with initial condition 
\be
\lim_{t\rightarrow0} K_{t}(\sigma)=\delta(\sigma,o),
\ee
where $\delta(\sigma,o)$ is the Dirac distribution on the spatial slice $\Sigma$ with support at $o$. (Note that, with the definition \eref{defMSD}, the MSD is measured as a function of the $t$ coordinate, which coincides with proper time only at the origin $o$: unlike the non-relativistic situation, there is no global physical time parameter in a curved spacetime.)


%

\subsection{Asymptotic expansion of the MSD}

Let us denote $\mathcal{D}$ the differential operator $\kappa\Delta_{q}(\chi\cdot)$. Then the equation \eref{staticdiffusion} can be solved formally as
\be
K_{t}(\sigma)=e^{t\mathcal{D}}\delta(\sigma,o)=\sum_{n=0}^{\infty}\f{t^{n}}{n!}\mathcal{D}^{n}\delta(\sigma,o). 
\ee
The MSD, in turn, can be computed by evaluating this distribution the squared distance function $\rho^{2}$. To this effect, note that
\be
\langle\mathcal{D}\delta,\rho^{2}\rangle=\langle\delta,\mathcal{D}^{\dagger}\rho^{2}\rangle=\mathcal{D}^{\dagger}\rho^{2}(o),
\ee
where $\mathcal{D}^{\dagger}=\kappa \chi\Delta_{q}$ is the formal adjoint of $\mathcal{D}$. Hence
\be\label{expansionMSD}
\langle\rho^{2}\rangle_{t}=\langle K_{t},\rho^{2}\rangle=\sum_{n=0}^{\infty}\f{t^{n}}{n!}(\mathcal{D}^{\dagger})^{n}\rho^{2}(o),
\ee
i.e. 
\be
\langle\rho^{2}\rangle_{t}=\sum_{n=0}^{\infty}\f{(\kappa t)^{n}}{n!}(\chi\Delta)^{n}\rho^{2}(o).
\ee
This formula provides the asymptotic expansion of the MSD in the small time limit $t\rightarrow0$. Up to second order in $t$, it gives 
\be
\langle\rho^{2}\rangle_{t}=2\kappa Dt\Big\{1+\Big(\f{\Delta\chi(o)}{2}-\f{R^{(D}(o)}{3D}\Big) \kappa t+\mathcal{O}\big(t^{2}\big)\Big\}.
\ee
To arrive at this expression we used the geometric identities $\Delta\rho^{2}(o)=2D$ and $\Delta^{2}\rho^{2}(o)=-4R^{(D)}(o)/3$.\footnote{The higher order terms involve higher derivatives of the squared distance function, which can also be expressed in terms of local curvature invariants \cite{DeWitt2003,Ottewill}.} At this order, we thus see that diffusion is enhanced by a convex lapse profile about $o$, and by negative spatial curvature.  

\subsection{The backward equation}

Note that the expansion \eref{expansionMSD} can be resummed formally as 
\be
\langle\rho^{2}\rangle_{t}=e^{t\mathcal{D}^{\dagger}}\rho^{2}(o). 
\ee
Thus, the MSD $\langle\rho^{2}\rangle_{t}$ can also be obtained as the solution $u_{t}$ to the adjoint, or \emph{backward}, equation
\be\label{backward}
\pp_{t}u_{t}=\kappa \chi\Delta u_{t}
\ee
with initial condition $u_{0}(\sigma)=\rho^{2}(\sigma)$. This differential formulation can be useful to obtain the MSD in concrete situations, by means of a numerical integration of \eref{backward}.

\subsection{Two examples}

We close this section with two explicit examples where the MSD is altered by the spacetime geometry. The first one is the simplest general-relativistic star model, and the second one is inspired by condensed-matter gravitational analogues such as graded-index optical fibers. 

\begin{itemize}
\item\emph{Schwartzschild's constant density star}. This is a static solution of the Einstein equation with uniform mass-energy density. It has two parameters $R$ and $M$, the radius and mass of the star respectively. (See \cite{Padmanabhan2010} for the explicit expression of the line element.) If $o$ is the center of the star ($r=0$), one computes $\Delta \chi(o)=3GM/R^{3}$ and $R^{3}(o)=12GM/R^{3}$, and therefore
\be
\langle\rho^{2}\rangle_{t}=6\kappa t\Big(1+\f{GM}{6R^{3}}\kappa t+\mathcal{O}(t^{2})\Big). 
\ee
Thus, a Brownian motion initialized at the center of the star spreads slightly \emph{faster} than in flat spacetime. This result might seem paradoxical: doesn't gravity \emph{attract}? Recall however that Brownian motion takes place within the stellar medium, which is not free-falling but static. The infinitely frequent collisions between this medium and the Brownian particle prevent the latter from falling down to the center of the star -- on the contrary, we see here that they actually \emph{increase} the MSD. Furthermore, this effect is small: a simple computation shows that the Brownian motion hits the surface of the star ($\langle\rho^{2}\rangle_{t}^{1/2}\simeq R$) long before the corrective term $(GM/6R^{3})\kappa t$ becomes of order $1$. 
\item\emph{Parabolic lapse profile}. Interestingly, this speed-up effect can be emulated, and amplified, in a gravitational analogue with flat spatial geometry and a parabolic lapse profile 
\be
\chi(\rho)=1+\epsilon \rho^{2}/R^{2}.
\ee 
Here $\epsilon=\pm1$ indexes the convexity/concavity of the profile. Such lapse profiles arise e.g. in graded-index optical fibers, or in Kerr media controlled by intense laser pulses. (In these optical contexts, the lapse function is nothing but the inverse of the refractive index.) Moreover, this case has the advantage that the asymptotic expansion \eref{expansionMSD} can be resummed explicitly. Indeed, we have
\be
\Delta \chi=\epsilon\Delta\rho^{2}/R^{2}=2\epsilon D/R^{2},
\ee 
hence the formula \eref{expansionMSD} gives 
\be
\langle\rho^{2}\rangle_{t}=2D\sum_{n=1}^{\infty}\f{\kappa^{n}t^{n}}{n!}\Big(\f{2\epsilon D}{R^{2}}\Big)^{n-1}
\ee
i.e. 
\be
\langle\rho^{2}\rangle_{t}=\epsilon R^{2}\Big(e^{2\epsilon D\kappa t/R^{2}}-1\Big).
\ee
In the convex case ($\epsilon=1$), the MSD therefore grows \emph{exponentially} with time, while in the concave case ($\epsilon=-1$), it slows down and eventually reaches the \emph{finite limit} $R^{2}$ on the ``infinite redshift surface'' $\{\chi=0\}$. This effect becomes significant on the time scale $R^{2}/\kappa$. Materials where $R$ can be tuned at will could therefore provide experimental benchmarks for the results discussed in this paper. 
\end{itemize}

\section{Conclusion}\label{conclusion}

Let us conclude. From a theoretical standpoint, our reasoning in this paper is very straightforward: it simply consists in incorporating gravitational redshift and spatial curvature effects into the standard master equation for a Markov process -- in short, Einstein 1905 \cite{Einstein1905} amended by Einstein 1912 \cite{Einstein1912a}. 

Simple as it is, though, this approach has allowed us to derive Eckart's constitutive relation for heat transfer, to generalize it to non-thermal diffusion processes, and to compute the gravitational correction to the diffusion square-root law. In particular, we have obtained the general asymptotic expansion of the mean-squared displacement in static spacetimes, and concluded from two explicit examples that experiments are more likely to reveal such corrections in \emph{analogue} gravitational systems. Given the ubiquity of diffusion phenomena in condensed-matter physics, we are hopeful that these results will prove useful in applications. This would confirm -- if that was needed -- that general relativity remains as fertile as ever.

\section*{Acknowledgements}

I am indebted to Eugenio Bianchi for a key conversation in Nice, in which we identified \eref{heat} as the general-relativistic heat equation. I also wish to thank the Marseille quantum gravity group, notably Carlo Rovelli, and Daniele Faccio for their constructive comments on this work, as well as the organizers of the SIGRAV school on analogue gravity (Como, May 2011), where I learnt about this rising field.



\section*{References}

\bibliographystyle{abbrv}
\bibliography{library}

\end{document}